\documentclass[twocolumn]{revtex4}
\usepackage[english]{babel}
\usepackage{amsmath}
\usepackage{amssymb}
\usepackage{graphicx}
\usepackage{rotating}

\begin{document}

\author{Ond{\v{r}}ej Stejskal}
\email[]{stejskal@karlov.mff.cuni.cz}
\affiliation{Institute of Physics, Charles University, Ke Karlovu 5, 12116 Prague, Czech Republic, \\
IT4Innovations, VSB-Technical University of Ostrava, 17.~listopadu 15, 70800 Ostrava, Czech Republic}

\author{Andr{\'{e}} Thiaville}
\email[]{andre.thiaville@universite-paris-saclay.fr}
\affiliation{Laboratoire de Physique des Solides, Universit{\'{e}} Paris-Saclay, CNRS UMR 8502, 91405 Orsay, France}

\author{Jaroslav Hamrle}
\affiliation{Institute of Physics, Charles University, Ke Karlovu 5, 12116 Prague, Czech Republic}

\author{Shunsuke Fukami}
\email[]{s-fukami@riec.tohoku.ac.jp}
\affiliation{Laboratory for Nanoelectronics and Spintronics, Research Institute of Electrical Communication, 
Tohoku University, \\
Center for Science and Innovation in Spintronics, Tohoku University, \\
Center for Spintronics Research Network, Tohoku University, \\
Center for Innovative Integrated Electronic Systems, Tohoku University, \\
WPI-Advanced Institute for Materials Research, Tohoku University, \\
2-1-1 Katahira, Aoba, Sendai, Miyagi 980-8577, Japan}

\author{Hideo Ohno}
\affiliation{Laboratory for Nanoelectronics and Spintronics, Research Institute of Electrical Communication, 
Tohoku University, \\
Center for Science and Innovation in Spintronics, Tohoku University, \\
Center for Spintronics Research Network, Tohoku University, \\
Center for Innovative Integrated Electronic Systems, Tohoku University, \\
WPI-Advanced Institute for Materials Research, Tohoku University, \\
2-1-1 Katahira, Aoba, Sendai, Miyagi 980-8577, Japan}

\title{Current distribution in metallic multilayers from resistance measurements}

\begin{abstract}

The in-plane current profile within multilayers of generic structure Ta/Pt/(CoNi)/Pt/Ta is investigated.
A large set of samples where the thickness of each layer is systematically varied was grown, followed by 
the measurement of the sheet resistance of each sample.
The data are analyzed by a series of increasingly elaborate models, from a
macroscopic engineering approach to mesoscopic transport theory.
Non-negligible variations of the estimated repartition of current between the layers are found.
The importance of having additional structural data is highlighted.
\end{abstract}

\maketitle

\section{Introduction}

From its very beginnings, spintronics is relying on metallic and semiconductor multilayers.
The samples where giant magnetoresistance (GMR) was discovered were indeed (Fe/Cr) multilayers \cite{Baibich88,Binasch89},
with current flowing in the layers' plane (CIP geometry).
The spin-transfer torque (STT) effect \cite{Berger84,Slonczewski96}, when used in the CIP geometry to drive magnetic
domain walls \cite{Grollier02,Klaui03b,Yamanouchi04,Yamaguchi04}, also involves metallic or semiconductor multilayers 
with magnetic and non-magnetic parts.
In such samples, only the total injected current is known, and its distribution between the various layers is an 
open question: how much of the current is diverted by the buffer layer, or by the spacer layer (case of spin-valve 
nanostrips \cite{Pizzini09}, or of synthetic antiferromagnet nanostrips \cite{Lepadatu17}) affects the
computed efficiency of the STT-induced domain wall motion.
The current in the non-magnetic layers also gives rise to a non-compensated Oersted field in the magnetic
layer, which affects the structure and dynamics of the domain walls \cite{Uhlir11}.
The spin-orbit torques (SOT), the latest family of current-induced torques in the CIP geometry 
\cite{Miron10,Liu11,Fukami16,Fukami16b}, also occur in such multilayer samples.
Here again, the distribution of the current between the various layers is of crucial importance.

Presently there exists no direct method to observe how current flows in the different layers of the
multilayer sample, hence indirect methods must be used.
The most employed one consists in using an effective conductivity for each layer, applying
the parallel conductors rule of macroscopic electrical engineering to evaluate how much of the current flows in
each layer.
This effective conductivity is either assumed, or in some cases obtained by growing a series of
samples with changing thickness of the layer(s) \cite{Fukami16b}, again using the rules of macroscopic electrical 
engineering to obtain this conductivity.

However, the increase of the apparent resistivity of a metal film when its thickness decreases below the 
mean free path of electrons has been discussed as early as 1901, by J.J. Thomson \cite{Thomson1901}.
The development of this idea has led to the so-called Fuchs-Sondheimer semi-classical model \cite{Fuchs38,Sondheimer52},
which has been further developed and refined.

In this paper, from an extensive series of metallic magnetic multilayer samples where all thicknesses
were varied, we quantitatively compare the samples conductivities with a number of models, with the goal of 
estimating the current distribution among the layers.
The samples are emblematic of present studies of STT and SOT causing, for example, current-induced domain wall motion.

\section{Samples description}

The stack structure of the multilayer reference sample used in this work is
Ta 3/ Pt 1.6/ [Co 0.3/ Ni 0.6]x4/ Co 0.3/ Pt 1.6/ Ta 3,
where thicknesses are given in nanometers.
The stacks were deposited by dc magnetron sputtering onto a thermally oxidized Si substrate.

\begin{figure*}
\includegraphics[width = 15 cm]{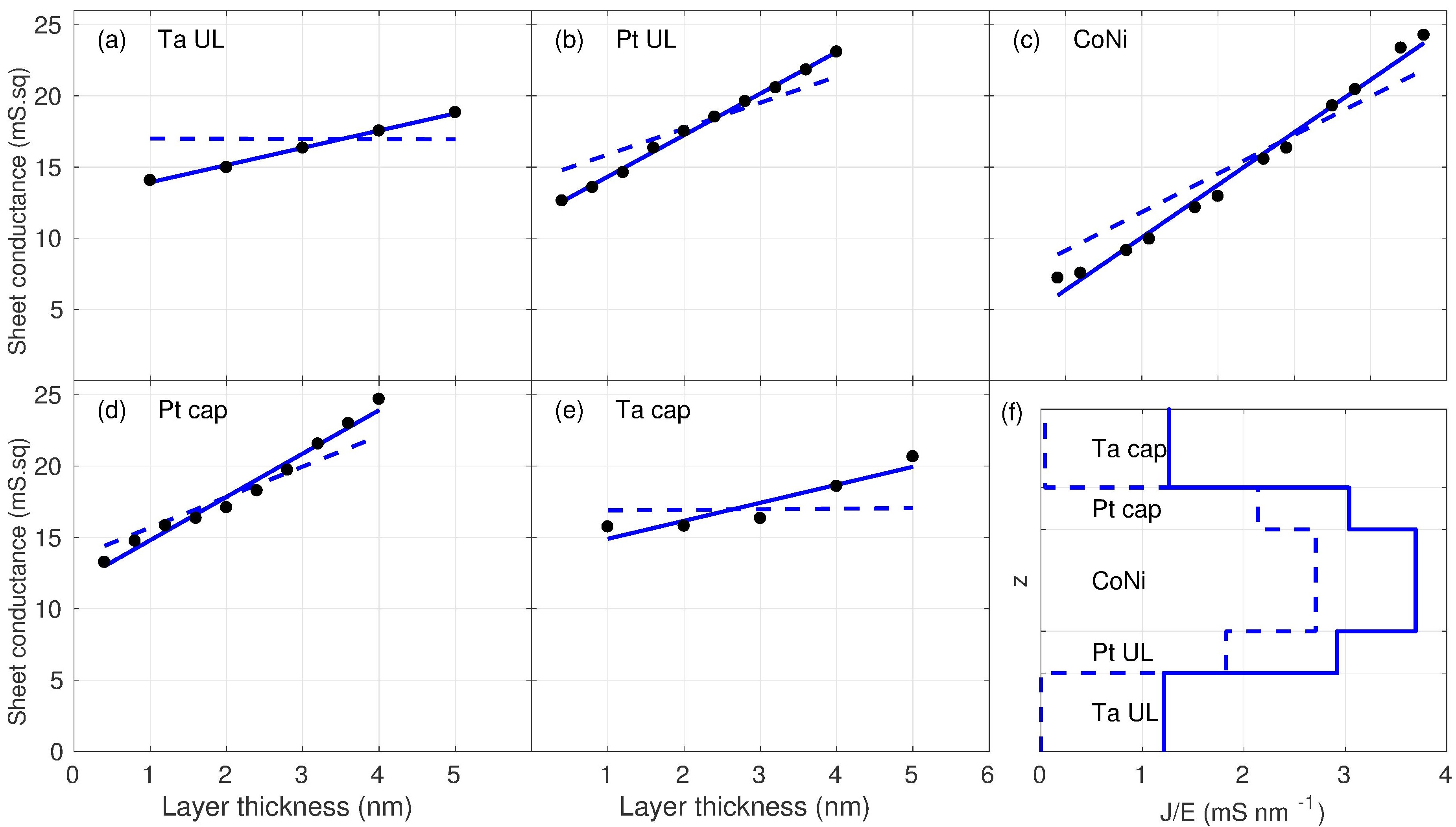}
\caption{
Measured sheet conductances (dots) vs. thickness of each layer (a-e), followed by the thickness-resolved
conductivity plot for the reference sample, as derived from the models (f).
The fit of each thickness series by a linear law (model 1) is represented by solid lines, whereas
the global fit according to Eq.~(\ref{eq:Glin}) (model 2) is drawn with dashed lines.
No structural information is taken into account here, so that the models used are called 1a resp. 2a.
The global rms error is 0.493~mS.sq when fitting each series independently, increasing to 1.621~mS.sq for 
the global fit.
In order to compare the slopes, all panels (a-e) have the same scales.
\label{fig:data+lin}
}
\end{figure*}

In order to estimate the current profile across the thickness, the thickness of each layer involved was
varied around its reference value, the other layers having their reference thickness.
This leads to a total of 42 samples, which were deposited consecutively with no change of the deposition parameters.
The sheet resistance of each sample was measured, just after deposition, by the four-probe technique, 
at 5 locations across the wafer, and the average of these measurements was taken.
The 5 values were extremely close (less than 1~{\%} difference), showing the uniformity of the film properties.
The reference sample was therefore fabricated 5 times. 
Slightly different resistance values were measured for this sample (from 58.2 to 64.3~$\Omega$/sq, with no trend 
in time), the average value being 61.2~$\Omega$/sq and the standard deviation 2.3~$\Omega$/sq, 3.7~\% of the mean value.
These variations can only be attributed to slow fluctuations of the deposition conditions, the noise with 
respect to the trend for each thickness series being at most 1~$\Omega$/sq.
To remove this noise in the raw data, the resistances of each thickness series were multiplied by a
series-specific factor such that the resistance of the reference sample becomes 61.2~$\Omega$/sq 
(using the raw data increases the error of all fits 
because of the data intrinsic noise, and trying to estimate the actual thicknesses is delicate).
The corresponding sheet conductances are shown in Fig.~\ref{fig:data+lin}.
We recall that, for a film of thickness $d$, the resistance of a strip of length $L$ and width $W$
is $R = \rho L / (W d) = (\rho / d) (L/W)$ with $\rho$ the apparent layer resistivity.
This introduces $R_\Box = \rho/d$ the sheet resistance, in Ohms per square ($\Omega$/sq), 
the aspect ratio of the strip $L/W$ being the number of squares of edge $W$ that can be inscribed in its
length $L$.
The sheet conductance $G_\Box$ is the inverse of the sheet resistance, with units Siemens.square (S.sq).
In order to give the same weight to each thickness series, the two Ta series with
only 5 data points were counted twice in the fitting.

\section{Independent layers model}

The first analysis of the data that can be performed is based on macroscopic electrical engineering,
where current flows in the different layers acting as parallel resistors.
Each layer (thickness $d_i$) is then described by a resistivity $\rho_i$ (conductivity $\sigma_i= 1/ \rho_i$),
its contribution to the sheet conductance being $\sigma_i d_i$.

Model 1 determines the $\sigma_i$ by a fit of the data, \emph{series by series}.
The straight solid lines in Fig.~\ref{fig:data+lin} show that this applies well to the data.
The fitted slopes correspond to resistivities 82.6, 79.3, 34.3, 32.9 and 27.1~$\mu \Omega$.cm for the 
tantalum underlayer, tantalum cap, platinum underlayer, platinum cap and CoNi multilayer, respectively.
The two tantalum resistivities are close, and of the expected order of magnitude for the high-resistivity
$\beta$ phase.
Those of platinum are close also, but about 3 times the pure metal value (10.7~$\mu \Omega$.cm).
The apparent resistivity of the (Co/Ni) multilayer is also about 4 times the values of cobalt and nickel.
At first sight, the resistivity values obtained from such an analysis may sound reasonable.
When computing the sheet conductance of the reference sample, however, one gets with these values
$G_\Box = 31.4$~mS.sq, close to twice the measured value (16.4~mS.sq).
Discarding the Ta thickness series results and assuming that no current flows in Ta is not enough to
solve the problem, as this leads to $G_\Box = 23.9$~mS.sq.

On the other hand, model 2 tries to fit \emph{globally} all data, according to
\begin{equation}
\label{eq:Glin}
G_\Box = \sum_i \sigma_i d_i.
\end{equation}
The results of this model are shown in Fig.~\ref{fig:data+lin} by the dashed lines.
Negative resistivities are obtained for the Ta underlayer thickness series
(see Tab.~\ref{tab:res-par}).
This again proves that one cannot treat the electrical conduction in such
multilayers made of nanometer-thick films by macroscopic engineering arguments, a fact known for a long time 
for single layers.
Indeed, the well-known Fuchs-Sondheimer model \cite{Fuchs38,Sondheimer52,Lucas65,Chopra69} with surface scattering 
leads to a reduction of current density close to the surfaces, on a scale given by the electron mean free path 
$\lambda$ (for example, with no specular scattering at both surfaces, the apparent conductivity of a single layer 
is reduced by a factor 2 for a thickness $d=0.46 \lambda$).

\section{Fuchs-Sondheimer model}
\label{FSm}

\begin{figure}
\includegraphics[width= 5 cm]{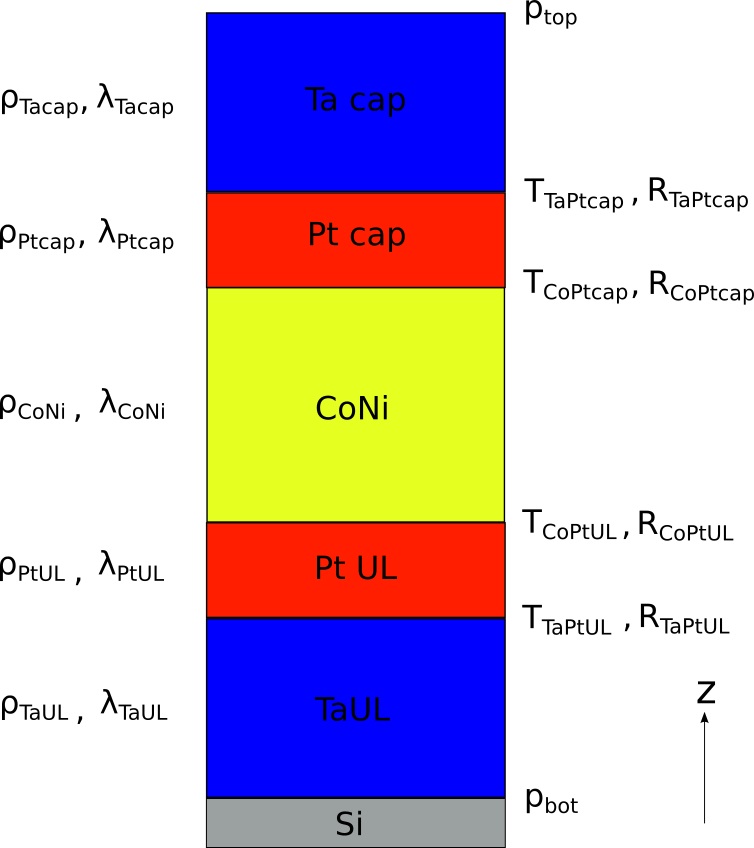}
\caption{
Scheme of the sample structure with the various parameters of the Fuchs-Sondheimer model applied to
analyze the conductivity data.
\label{fig:scheme}
}
\end{figure}

As the samples are multilayers, we use the multilayer formalism described by 
Barna{\'{s}} {\it{et al.}}~\cite{Barnas90} of the Fuchs-Sondheimer (FS) model.

The detailed derivation of the model can be found in the aforementioned publications, so here we present only the 
resultant expressions. 
The distribution function for electrons in a given layer can be written in the form:
\begin{equation}
f(z,\mathbf{v})=f_0(\mathbf{v})+g(z,\mathbf{v}),
\label{eq:f1}
\end{equation}
where $f_0(\mathbf{v})$ is the equilibrium distribution function and $g(z,\mathbf{v})$ corresponds to the contribution 
induced by the external electric field. 
The $z$-direction is perpendicular to the interfaces (Fig.~\ref{fig:scheme}). 
By solving the Boltzmann equation, one obtains the general expression for $g_\pm(z,\mathbf{v})$  ($g_+$ corresponding 
to the electrons moving in the positive $z$ direction, $g_-$ to the electrons moving in the negative $z$ direction),
$d$ being the layer thickness:
\begin{eqnarray}
g_+(z,\mathbf{v}) &=& \frac{eE\tau}{m}\frac{\partial f_0 (\mathbf{v})}
{\partial v_x}\left[ 1-F_+(\mathbf{v}) \exp \left( \frac{- z}{\tau |v_z|} \right) \right], \\
g_-(z,\mathbf{v}) &=& \frac{eE\tau}{m}\frac{\partial f_0 (\mathbf{v})}
{\partial v_x}\left[ 1-F_-(\mathbf{v}) \exp \left( \frac{-(d-z)}{\tau |v_z|} \right) \right], \nonumber
\label{eq:g1}
\end{eqnarray}
where $F_\pm(\mathbf{v})$ are functions to be determined from the boundary conditions. 
The symbols $e$ ($e>0$) and $m$ are the electron charge and effective mass, $\tau$ is the relaxation time, 
$v_x$ and $v_z$ the components of the velocity vector in $x$- and $z$-directions, respectively. 

\begin{figure*}
\includegraphics[width = 15 cm]{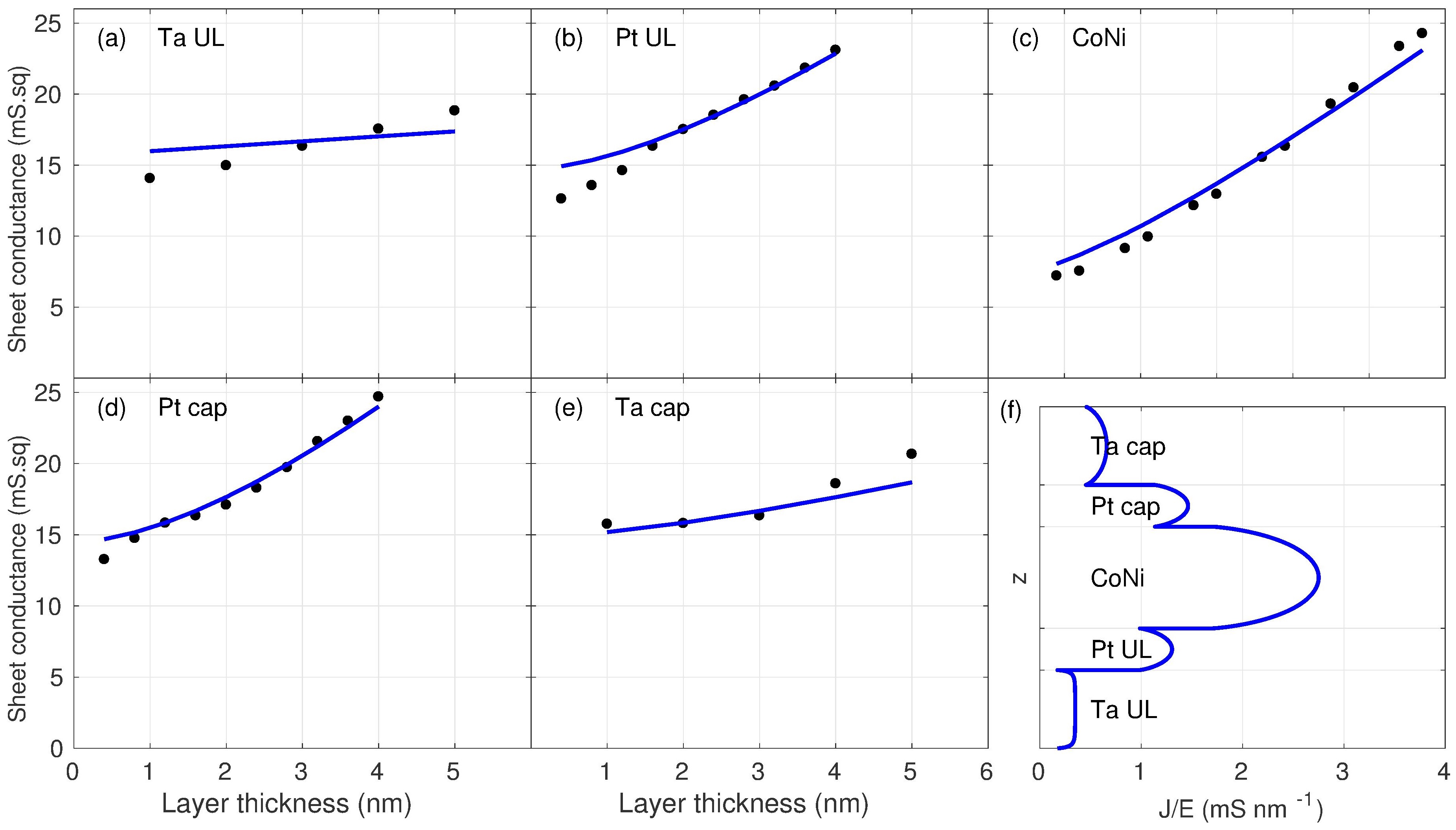}
\caption{
Fuchs-Sondheimer fit of all the data, with assumptions described in the text, and without any input 
from the structural characterization. 
The resulting rms error is 0.978~mS.sq (model 3a). 
\label{fig:FS1}
}
\end{figure*}
\begin{figure*}
\includegraphics[width = 15 cm]{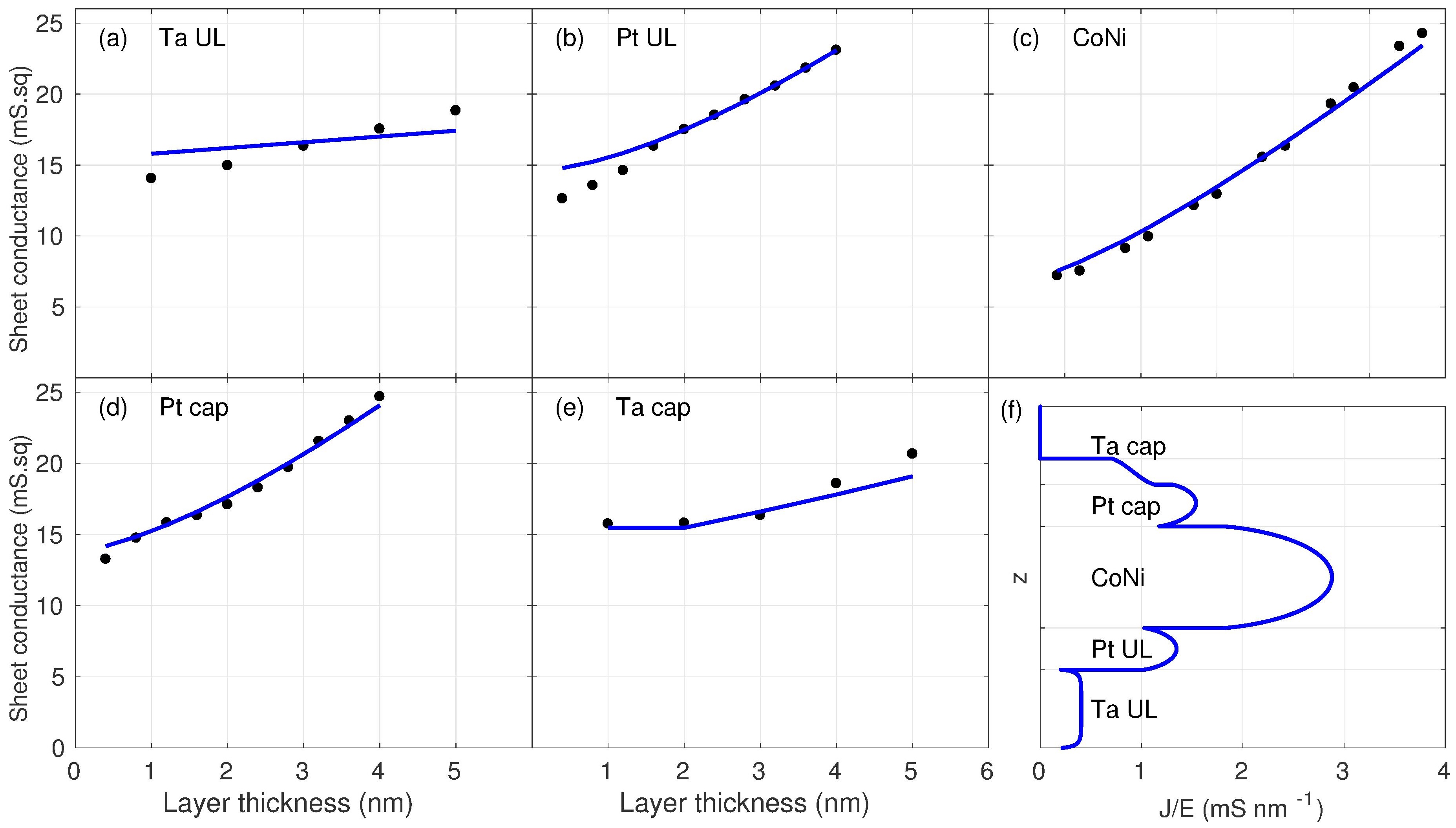}
\caption{
Fuchs-Sondheimer fit of all the data, taking into account a 2.0~nm oxidized Ta cap thickness. 
The corresponding rms error is 0.829~mS.sq (model 3b).
\label{fig:FS2}
}
\end{figure*}

\subsection{Direct application to the data}
\label{FSm1}

We impose the Fuchs boundary conditions at the bottom and top interface using coefficients $p_\mathrm{bot}$ and 
$p_\mathrm{top}$ corresponding to the specularity factors \cite{Lucas65}. 
For the inner interfaces, similarly to Ref.~\onlinecite{Barnas90}, we introduce coefficients of specular transmission 
$\mathrm{T}$ and reflection $\mathrm{R}$
which differ for each interface (Fig.~\ref{fig:scheme}), but we neglect any angular dependence of these 
coefficients, an approximation at the heart of the FS model. 
We also neglect the refraction effects that may occur when the Fermi velocities in adjacent materials are different, 
and we assume the same transmission and reflection coefficients for electrons incident on the interface from the top 
and bottom. 
For simplicity, we also neglect any spin dependency of the coefficients, no magnetic field effects being investigated.
For the (Co/Ni) multilayer, given the small thickness of the individual layers, the atomic proximity of
Co and Ni, their good alloying properties, and the observed weak variation of resistivity with alloy concentration
\cite{Muth81}, it is treated as a single layer (alloy).

Application of these boundary conditions leads to a set of $N$ equations of $N$ variables, where $N$ is twice the number 
of layers in the sample (for every layer we have $F_\pm$). 
These equations are the spin-independent versions of the ones shown in Ref.~\cite{Barnas90}. 
We introduce $\beta$, the angle between the $z$ axis and the velocity vector $\mathbf{v}$, and $\lambda=v_\mathrm{F} \tau$
the electron mean free path, $v_\mathrm{F}$ being the Fermi velocity. 
Note that all mentioned parameters [$\lambda, v_\mathrm{F}, \tau, m$] will generally differ for each 
layer. 
In correspondence with \cite{Barnas90}, we define 
$y_\mathrm{AB}=\frac{m_\mathrm{A} \tau_\mathrm{B}}{m_\mathrm{B} \tau_\mathrm{A}}$ to take into account 
the difference in electronic properties of two adjacent materials A and B. 
In our calculations we set $y_\mathrm{AB}=\sigma_\mathrm{B}/\sigma_\mathrm{A}$, where 
$\sigma_\mathrm{A}$ ($\sigma_\mathrm{B}$) is the bulk conductivity of material A (B). 
These equations are solved numerically for each $\beta$ to obtain $F_\pm(\beta)$.

The final formula for the apparent conductivity of a single layer is:
\begin{widetext}
\begin{equation}
\sigma_i=\sigma_{0,i}-\frac{3}{4}\frac{\sigma_{0,i}}{d_i}\lambda_i \int_0^{\pi/2} \mathrm{d}\beta \,\,
\sin^3 \beta \cos\beta
\left[F_{i,+}(\beta)+F_{i,-}(\beta)\right]\left(1-\exp\left(-\frac{d_i}{\lambda_i \cos\beta}\right)\right),
\end{equation}
\end{widetext}
where $\sigma_{0,i}$ and $d_i$ are the bulk conductivity and thickness of layer number $i$, respectively. 
By Eq.(~\ref{eq:Glin}), we obtain the sheet conductance of the multilayer sample, which is compared to the 
experimental results.

The total number of parameters entering the fit is 20 (Fig.~\ref{fig:scheme}), and it is further reduced. 
First, following Barna{\'{s}} \cite{Barnas90}, we neglect reflection at the inner boundaries, the work functions being
very close for all metal layers. 
Second, we fix the $\rho \lambda$ product to the reference values for each layer \cite{Vancea84,Stella09}.
The values adopted for the $\rho \lambda$ product (in $f \Omega$.m$^2$) were, from bottom (Ta underlayer) to top (Ta cap)
0.74, 2.35, 1.26, 2.35, and 6.0; see next section for justification of the choices for the Ta values.
Additionally, upper boundaries for the mean-free path $\lambda$, equivalent to lower boundaries for the resistivity $\rho$
given by the best samples values, were set at 0.3, 22, 18, 22, and 10 nm, respectively.
Third, we set $p_\mathrm{bot}=p_\mathrm{top}=0$ on the outer boundaries, as they correspond to layers carrying little 
current so that the fit error depends weakly on them.
After this reduction we are left with 9 parameters for all 42 samples. 
We fit the mean free paths of all the layers and transmission coefficients at all four inner interfaces.

The best fit of all data by the FS model (model 3) is shown in Fig.~\ref{fig:FS1}, together with the 
corresponding computed current profile across the thickness, for the reference sample. 
The rms error of this fit is 0.978~mS.sq; it is larger than that of the non-physical fit of each thickness series 
(model 1), but much lower than the error of the engineering-like independent layers model (model 2).
The obtained parameters are given in Tab.~\ref{tab:res-par}, and the current partition between the different layers
as computed by the various models are compared in Tab.~\ref{tab:res-Ji}.

\begin{figure*}
\includegraphics[width = 15 cm]{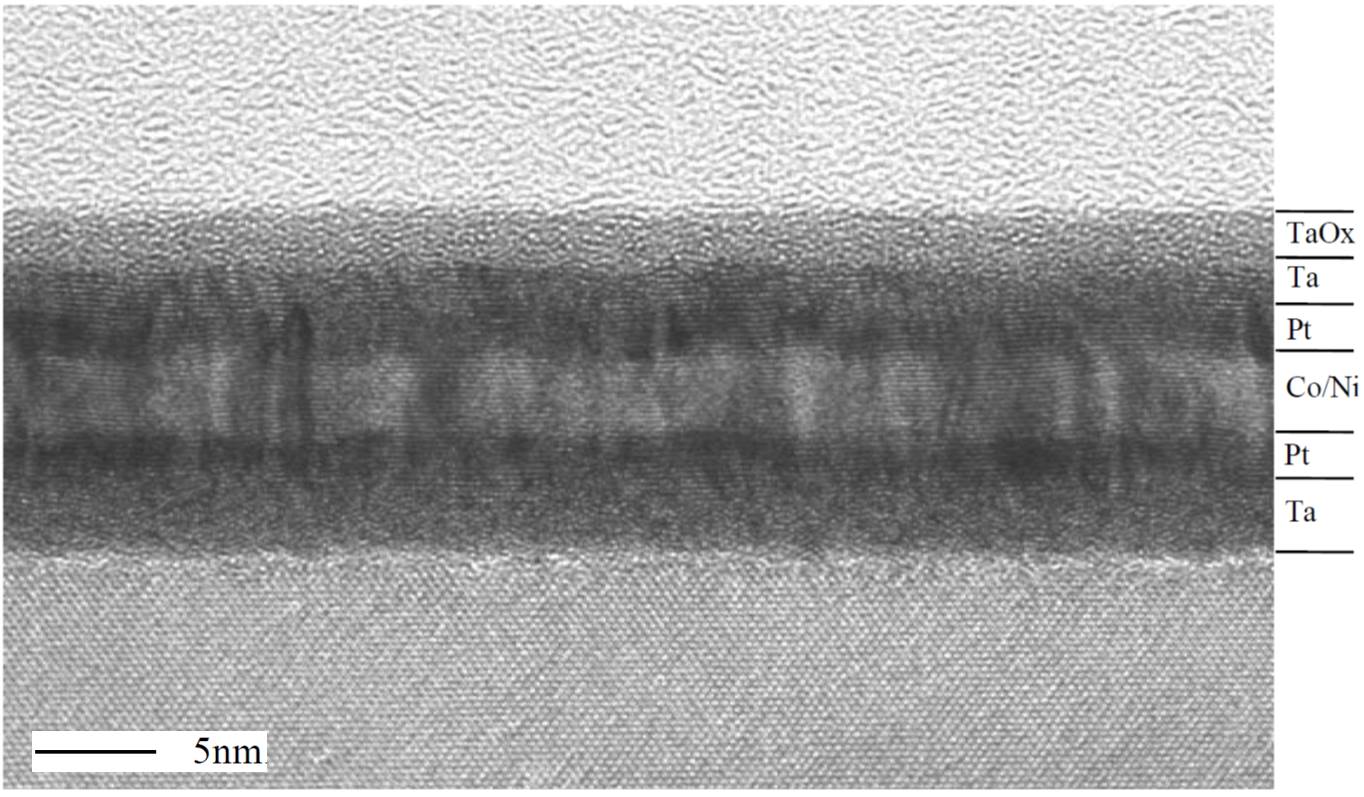}
\caption{
High magnification cross-sectional TEM image of the reference sample.
\label{fig:TEM-cut}
}
\end{figure*}
\begin{figure*}
\includegraphics[width = 10 cm]{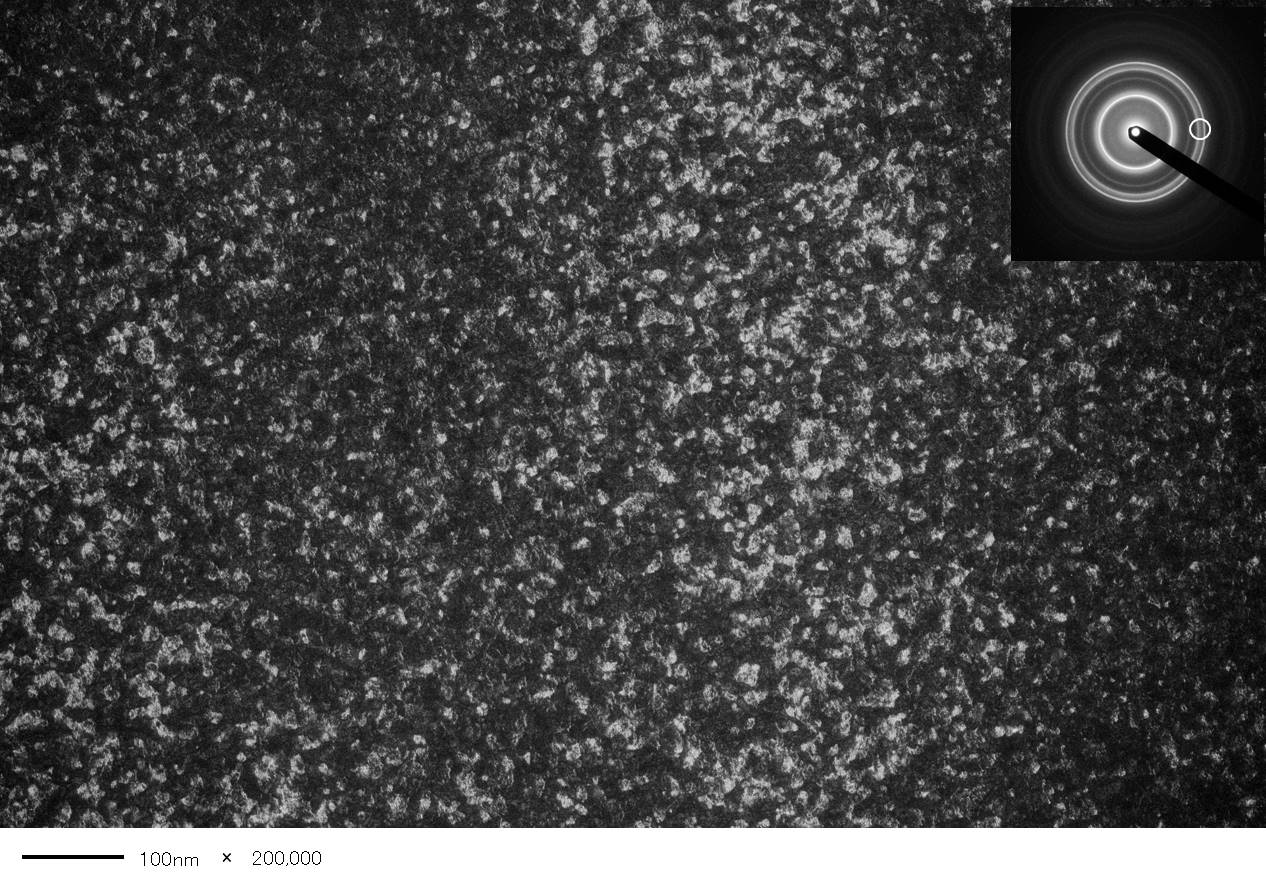}
\caption{
Dark-field TEM plane view of the reference sample, obtained by filtering out the (220) diffraction for
Pt, Co and Ni (see diffraction pattern in inset).
\label{fig:TEM-plane}
}
\end{figure*}

\subsection{Use of structural data}
\label{FSm2}

On close inspection, the acquired data appear strange at two instances.
First, in the Ta cap series, one notices that the first three points show 
nearly the same conductivities.
A cross-sectional transmission electron microscopy (TEM) image of the reference sample, shown in Fig.~\ref{fig:TEM-cut}, 
reveals that the Ta cap layer is partly oxidized, with about 2~nm Ta consumed.
Thus, the various models' fits were performed again under this assumption (see Tab.~\ref{tab:res-par} for the 
corresponding data).
This variant of the models is indicated by the suffix `b', suffix `a' indicating that the models are 
applied to the data with nominal thicknesses.
For the independent layers model, the rms error decreased to 0.423~mS.sq when fitting each series separately, whereas
in the case of the global fit the relative decrease was smaller, the error reaching 1.555~mS.sq.
For the FS model, as shown in Fig.~\ref{fig:FS2}, the quality of the fit improves markedly.

The TEM image also reveals that the crystalline structures of the two Ta layers are different, the underlayer
showing no sign of crystallinity whereas the cap layer does.
This is reflected in the fitted values of the FS model with the product $\rho \lambda$ fixed according to
the literature values for the two phases of tantalum \cite{Stella09}.

Second, the platinum underlayer series, when compared to the FS fit, [Fig.~\ref{fig:FS2}(b)] shows a break for the 
three lowest thicknesses (0.4-1.2~nm).
This is tentatively attributed to the formation of textured crystallites as the Pt underlayer starts growing on
the amorphous Ta underlayer.
These crystallites were directly imaged in TEM (Fig.~\ref{fig:TEM-plane}) for the reference sample, giving a typical
lateral size of 10~nm.
The next section describes the modeling of the additional electron scattering at the grain boundaries.

\begin{table*}[t]
\caption{Models used to fit the resistance data, their rms errors and parameters (layers
are numbered from bottom to top, so 1 is the Ta underlayer etc.).
Model 1 considers independent layers, fitting each thickness series separately.
Model 2 also considers independent layers, but fits all thickness series together.
Model 3 is the Fuchs-Sondheimer model.
Model 4 is the Mayadas-Shatzkes model, with 2.0~nm oxidized Ta cap, the grain size varying in proportion 
to the fcc layers thickness (model 4), or to the Pt underlayer thickness (model 4*). 
Suffix `a' indicates that the nominal thicknesses are considered, whereas suffix `b' means that up to
2.0~nm of Ta cap was considered oxidized.}
\label{tab:res-par}
\begin{center}

\begin{tabular}{ c | c | c c c c c | c c c c c | c c c c | c }
\hline \hline 
Model &  error &  $\rho_1$  & $\rho_2$ & $\rho_3$ & $\rho_4$ & $\rho_5$ &
 $\lambda_1$  & $\lambda_2$ & $\lambda_3$ & $\lambda_4$ & $\lambda_5$ &
 $T_{12}$ & $T_{23}$ & $T_{34}$ & $T_{45}$ & $R_\mathrm{grain}$ \\  \
   & (mS.sq) & \multicolumn{5}{c|}{($\mu \Omega$.cm)}  & \multicolumn{5}{c|}{(nm)}
 & \multicolumn{4}{c|}{()} & ()  \\ \hline 

1a & 0.49 & 82.6 & 34.3 & 27.1 & 32.9 & 79.3 &  &  &  &  &  &  &  &  &  &  \\

1b & 0.42 & 82.6 & 34.3 & 27.1 & 32.9 & 62.2 &  &  &  &  &  &  &  &  &  &  \\
\hline

2a & 1.62 & -8300  & 54.8 & 36.9 & 46.8 & 2590 &  &  &  &  &  &  &  &  &  &  \\

2b & 1.56 & -807  & 56.8 & 38.3 & 48.2 & 101 &  &  &  &  &  &  &  &  &  &  \\
\hline

3a & 0.98 & 287 & 22.7 & 25.2 & 16.4 & 75.5 & 0.26 & 10.3 & 5.00 & 14.4 & 7.95 & 0 & 0 & 0 & 0 &  \\

3b & 0.83 & 247 & 20.8 & 23.3 & 18.5 & 60.0 & 0.3 & 11.3 & 5.40 & 12.7 & 10.0 & 0 & 0 & 0 & 0.78 &  \\
\hline

4b & 0.71 & 247 & 31.1 & 18.0 & 11.8 & 60.0 & 0.3 & 7.6 & 7.0 & 19.9 & 10.0 & 0 & 0 & 0 & 0.97 & 0.19 \\

4*b & 0.67 & 247 & 29.7 & 12.1 & 14.1 & 60.0 & 0.3 & 7.9 & 10.4 & 16.7 & 10.0 & 0.24 & 1.0 & 1.0 & 1.0 & 0.57 \\

\hline
\hline
\end{tabular}
\end{center}
\end{table*}
%

\section{Extended Fuchs-Sondheimer model}
\label{EFSm}

The FS model alone is not sufficient for the reproduction of experimental data. 
One reason for that is the polycrystalline structure of the layers in the sample, revealed e.g. by TEM imaging
both in transverse cut (Fig.~\ref{fig:TEM-cut}) and in plane view (Fig.~\ref{fig:TEM-plane}). 
The boundaries between crystallites introduce an additional scattering of the electrons that is not covered by the FS model. 
This phenomenon was addressed by Mayadas and Shatzkes~\cite{Mayadas70}.

The FS model also does not account for roughness and thickness fluctuation of the layers, that naturally occur 
in the samples.
Several models have been proposed to account for these phenomena; this will be studied in a second part.

\subsection{Mayadas-Shatzkes model}

\begin{figure*}
\includegraphics[width = 15 cm]{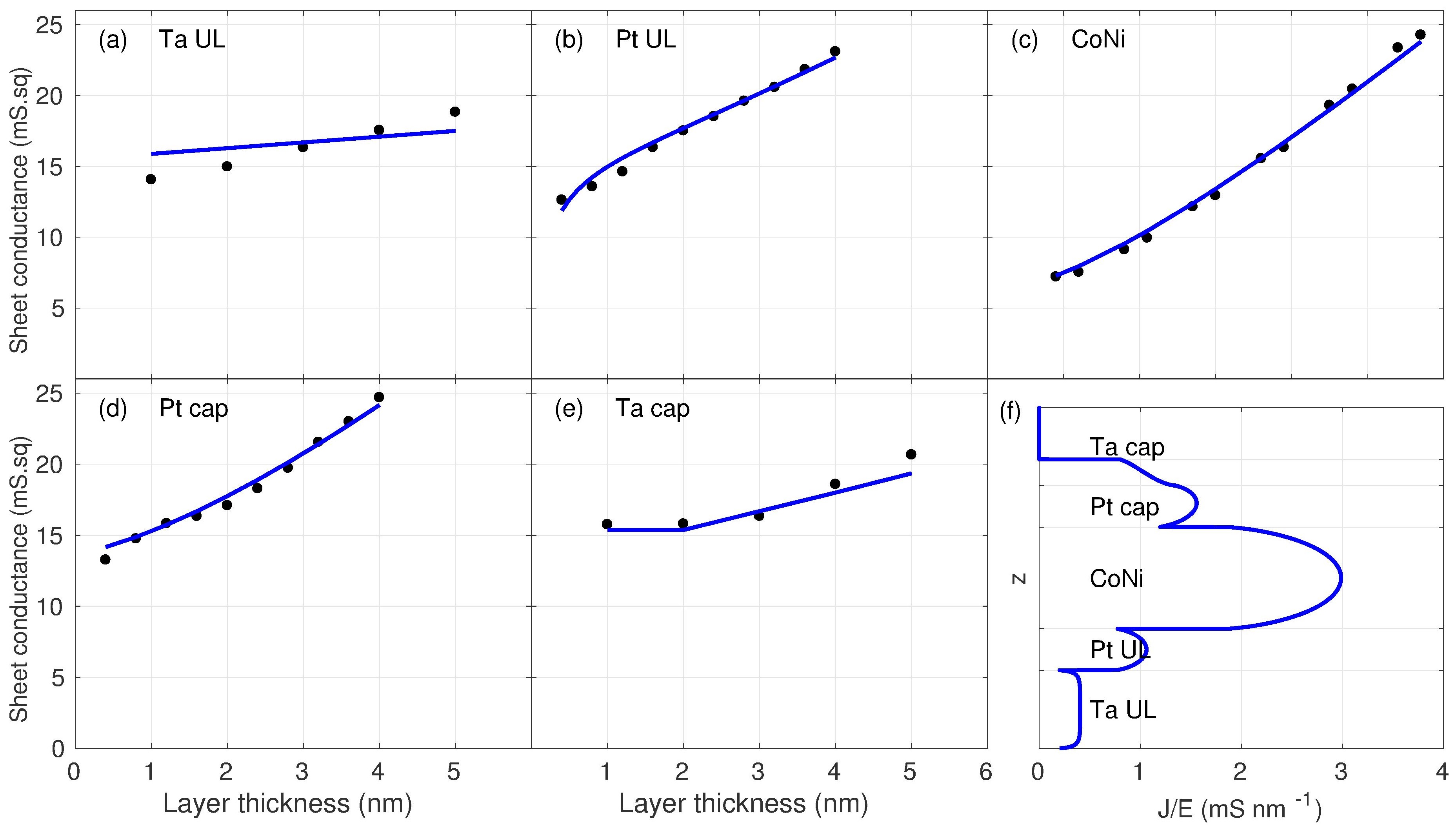}
\caption{
Mayadas-Shatzkes fit, with assumptions described in the text, taking into account that 2.0~nm of the
Ta cap were oxidized, the grain size varying in proportion of the Pt underlayer thickness (model 4b).
The resulting rms error is 0.705~mS.sq.
The curve downwards bend at low platinum thickness in panel (b) directly depicts the assumption of model 4;
it is absent in the other variant (model 4*, not shown).
\label{fig:FSMS2}
}
\end{figure*}

The model by Mayadas {\it{et al.}}~\cite{Mayadas69, Mayadas70} introduces grain boundary scattering into the FS model. 
For bulk samples or thick films, grain boundaries have little effect on the resistivity of metals as the grain 
size is much larger than the mean free path, however this is not the case for thin films, where the distance between 
grain boundaries is much smaller.

The modification of the FS model to account for the grain boundary scattering is simple to implement 
(for full derivation see Ref.~\cite{Mayadas70}), the relaxation time being redefined to
\begin{equation}
\frac{1}{\tau^*}=\frac{1}{\tau} \left(1+\frac{\alpha v_\mathrm{F}}{|v_x|} \right)
=\frac{1}{\tau} \left( 1+\frac{\alpha}{|\sin\beta\cos\phi|} \right)
:=\frac{M(\beta,\phi)}{\tau},
\end{equation}
where $\phi$ is the angle between the $x$ axis and the projection of the velocity vector $\mathbf{v}$ into the $xy$ plane. 
The parameter $\alpha$ carries the information about the grains and is defined as
\begin{equation}
\alpha=\frac{\lambda}{D}\frac{R_\mathrm{grain}}{1-R_\mathrm{grain}},
\end{equation}
with $D$ the typical grain diameter and $R_\mathrm{grain}$ the grain boundary reflection coefficient.

The final formula for conductivity of a single polycrystalline layer thus reads
\begin{widetext}
\begin{equation}
\sigma = \frac{3}{\pi}\frac{\sigma_0}{d} \int_0^{\pi/2} \int_0^{\pi/2} \mathrm{d}\phi \, \mathrm{d}\beta \,\,   
\frac{\cos^2\phi \sin^3\beta}{M(\beta,\phi)} \bigg\{ 2d 
- \frac{\lambda\cos\beta}{M(\beta,\phi)}
\left[F_{+}(\beta,\phi)+F_{-}(\beta,\phi)\right] 
\left(1-\exp\left(-\frac{d}{\lambda} \frac{M(\beta,\phi)}{\cos\beta} \right)\right)  \bigg\}.
\end{equation}
\end{widetext}

\begin{table}[t]
\caption{Current distribution in the sample calculated by the various models ($I_n$ is
the proportion of the total current that flows in layer $n$, in percent). 
See Tab.~\ref{tab:res-par} for the definition of models, their fitted parameters, and the 
residual error of each fit.
Model 0 is the uniform distribution of current throughout the sample, the variant (model 0*)
assuming that no current flows in the tantalum, because of its very high resistivity
in the amorphous phase.}
\label{tab:res-Ji}
\begin{center}
\begin{tabular}{ c | c c c c c | c c c c c }
\hline 
Model & \multicolumn{5}{c|}{nominal thicknesses} & \multicolumn{5}{c}{w/ oxidized Ta cap} \\
      & \multicolumn{5}{c|}{ (models `a')} & \multicolumn{5}{c}{ (models `b')} \\
 & $I_1$ & $I_2$ & $I_3$ & $I_4$ & $I_5$ & $I_1$ & $I_2$ & $I_3$ & $I_4$ & $I_5$ \\  
 \hline 

0 & 23 & 12 & 30 & 12 & 23    & 27 & 14 & 35 & 14 &  9 \\

0* & 0 & 23 & 55 & 23 & 0     & 0 & 23 & 55 & 23 & 0  \\

1 & 12 & 15 & 46 & 16 & 12    & 12 & 16 & 49 & 17 &  6 \\

2 &  -0.2 & 17 & 62 & 20 & 0.7  & -2 & 17 & 60 & 20 &  6  \\

3 &  6 & 12 & 58 & 13 & 11    &  7 & 12 & 61 & 14 &  6 \\

4 &  &  &  &  &               & 7 & 9 & 63 & 14 &  6\\

4* &  &  &  &  &                & 7 & 14 & 56 & 16 &  7\\

\hline
\end{tabular}
\end{center}
\end{table}

For the fits, the $\rho\lambda$ products and zero reflectivities are kept as before to make results comparable, 
but $R_\mathrm{grain}$ is included as an additional fit parameter, resulting in 10 free parameters. 
The cross-sectional TEM image (Fig.~\ref{fig:TEM-cut}) gives information on the grain structure.
The Ta underlayer is found amorphous, thus no grains are considered in the model (technically, 
$R_\mathrm{grain}=0$ is assumed for the Ta underlayer). 
For the Ta cap layer, results are weakly sensitive to grains, and the cross-sectional TEM image does not reveal
them.
No grain effect is therefore assumed for this layer ($R_\mathrm{grain}=0$).
For the two platinum and the (Co/Ni) layers, that all grow with an fcc (111) texture (Fig.~\ref{fig:TEM-plane}), 
cylindrical grains are considered, with the same reflection coefficient $R_\mathrm{grain}$.

In the absence of systematic TEM images, the variation of the size of the grains as a function of the thickness 
of these three layers has to be modeled.
The following results are based on speculations about the grain sizes.
A study having observed that platinum grains grow linearly with thickness below 50~nm \cite{Melo04}, two models
(at least) can be considered: the grain size is proportional to the Pt underlayer thickness (model 4), or
to the total fcc stack thickness (model 4*).
In both cases, the proportionality constant is such that, for the reference thickness, the grain size matches
the observed value of 10~nm.

Figure~\ref{fig:FSMS2} shows the results for model 4.
The obtained value $R_\mathrm{grain} \approx 0.2$ is typical \cite{Mayadas70}
(a variant of model 4 was also tried, in which $R_\mathrm{grain}=0.2$ was fixed;
the resulting rms errors are 0.705 and 0.755~mS.sq for the models `b' and `*b', respectively).
A clear improvement is obtained relatively to the FS model, the current partition between the layers being 
slightly affected (see Tab.~\ref{tab:res-Ji}).
The evolution of the fitted intrinsic parameters for the platinum underlayer is surprising, and it differs
from what is seen for the other fcc layers.
As the fit is still not fully satisfactory (see the tantalum underlayer data), a parasitic effect of the fit,
which amounts to forcing data to comply to a given framework, cannot be ruled out. 
Such an effect was already observed for model 2, where negative resistivity and current for the
tantalum underlayer were obtained.
One should also not forget that, as more extrinsic scattering sources are added (the interfaces are added
in model 3, the grains are added in model 4), the current distribution becomes less dependent
on the intrinsic parameters $\rho_i$ and $\lambda_i$.

The alternative grain size hypothesis (model 4*) delivers a slightly better fit, and redistributes the 
current more evenly among the fcc layers, due to the fact that the fitted internal specularity factors 
have reached unity. 
In addition, the grain reflection coefficient is found much larger.
This sensitivity of the fitting might just signal the too large number of free parameters.
Globally however, models 4 and 4* predict similar current partitions.

\subsection{Thickness averaging}

All the above modeling was assuming layers that are perfect in the $z$ direction, despite the common observation that
some sample roughness exists and that the layer thicknesses fluctuate from place to place (see for example the
cross-sectional TEM image in Fig.~\ref{fig:TEM-cut}).
In the literature, it has been proposed to account for this fact either by modifying the specularity parameter 
\cite{Brandli65,Parrott65,Soffer67}, or by averaging over a thickness distribution \cite{Namba70}.
The latter model, considering only one-dimensional thickness fluctuations, is clearly oversimplified.
Two-dimensional fluctuations in conductor networks have been deeply studied more recently, and an effective medium 
approximation (EMA) has been shown to describe well the regime beyond the percolation transition \cite{Luck91}. 
For averaging in two dimensions, the EMA formula is an implicit equation:
\begin{equation}
\left\langle \frac{\sigma-\sigma^{\mathrm{EMA}}}{\sigma+\sigma^{\mathrm{EMA}}}\right\rangle=0,
\label{eq:ema}
\end{equation}
where the brackets denote the averaging over the probability distribution $\rho(\sigma)\mathrm{d}\sigma$. 

This effect has been implemented for all models, assuming that the fcc grains, being slightly
disoriented as proved by the diffraction image Fig.~\ref{fig:TEM-plane}, can grow at different speeds
with the largest speed when the (111) plane lies perfectly horizontal.
To model this, a normal distribution of the growth rate was taken, with a standard deviation expressed as a
percentage $p$ as variable parameter.
This corresponds to recent detailed studies in magnetic ultrathin films of skyrmions morphology
and pinning \cite{Gross18},  and of domain wall dynamics \cite{Hrabec18}.
Typical values of $p$ extracted from these works are $p= 1-3$~\%.
To compute the average in Eq.~(\ref{eq:ema}), for each data point, the model values for all the fcc thicknesses
modified according to the same value of the normal probability law were used for integration, repeating the 
average calculation for 10 values of $\sigma^\mathrm{EMA}$ close to the model value in order to interpolate
the solution of the EMA equation.

It was found that the EMA refinement led to negligible fitting error improvements, and thus negligible
changes of the fitting parameters, even if values of $p$ as large as 5~\% were considered.
The mathematical reason for this is simple: as in Eq.~(\ref{eq:ema}) the low conductances get a larger
weight than the large conductances, the bias of the EMA averaging is towards lower conductances,
with shifts that increase as the thickness variation is larger.
This results in a systematic negative curvature of the conductance versus thickness curves, quite the
opposite of the experimental trend.

\section{Conclusion}

The schematic types of models for the current distribution are: uniform current through the conductive
layers (model 0), constant conductivity for each layer, obtained from the corresponding
thickness series (model 1), and Fuchs-Sondheimer global model(model 3).
We have shown that increasingly better fits of the data are obtained by considering more and more
elaborate models (designated by higher and higher model numbers), taking into account additional
structural data.

The remarks emerging from this work are.
\begin{itemize}
\item{Comparing model 1 to 3, we see that the former underestimates the current in the CoNi
multilayer, overestimates it in the Pt layers, and largely overestimates it for the
Ta layers. 
This was to be expected, due to mean-free path effects.}
\item{Model 2 is too rarely employed, which is a pity as this model shows directly 
the danger of model 1.}
\item{The Fuchs-Sondheimer model (model 3) provides results that are not bracketed by the 
simplest models (models 0 and 1), as might have been assumed. 
These models should therefore be taken as indicative only.}
\item{The Mayadas-Shatzkes model (model 4) does improve the fit error, but does not
modify much the current distribution.}
\end{itemize}
On this basis, the simplest way to improve model 1 seems to perform a Fuchs-Sondheimer
calculation with zero specularity coefficients at the interfaces, using reasonable
values for the intrinsic parameters.

The work reported shows that collecting as much as possible structural data is desirable.
From breaks of slope in a thickness series one might guess that another phenomenon is taking place,
but it is always better to have an independent observation of it.
The effect of including structural information is not negligible, compare for example the current in the 
Ta cap layer as deduced from models 3a and 3b.

Concerning non-uniformities, which are often invoked, we have found that for these samples
at least they play a negligible role.
This conclusion was reached using an analytical averaging formula, simpler that
numerical models proposed previously \cite{Elsom81}.
This conclusion opposes the one reached some time ago by Hoffmann et al. \cite{Hoffmann81},
probably due to the better quality of the films prepared nowadays.

One feature of the data that remains unclear is the influence of the Ta underlayer
thickness on the sample conductivity.
The observed increase of sheet conductance with thickness of this layer is indeed too large,
so that model 1 predicts a large percentage of current in that layer.
One may suspect that this is a texture effect.
The role of a Ta underlayer on the subsequent growth of Pt, followed by Co, has been indeed recently
highlighted by measurements of the interfacial Dzyaloshinskii-Moriya interaction \cite{Kim15}, an
anti-symmetric exchange term arising at the Co/Pt interface, which is anticipated to depend strongly 
on the interface quality \cite{Yang15}.
To test this hypothesis, TEM studies of the sample as the Ta underlayer thickness is varied
would be decisive.

\begin{acknowledgments}

This work was partly supported by Czech Science Foundation (project 19-13310S), by the European 
Regional Development Fund in the IT4Innovations national superconducting center - path to the exascale
(project CZ.02.1.01/0.0/0.0/16{\_}013/0001791), by the japanese FIRST Program of JSPS and Kakenhi 
Grant Number 19H05622.

\end{acknowledgments}


\end{document}